\documentclass[runningheads]{llncs}

\usepackage[vlined, linesnumbered, noend]{algorithm2e}
\SetAlCapSkip{1em} 

\usepackage{mathtools} 
\usepackage{amsfonts} 
\usepackage{amsmath} 
\usepackage{array,booktabs}
\usepackage{listings}
\usepackage{tikz}
\usepackage{float}
\usepackage{multirow}
\usepackage{todonotes}
\usepackage{lipsum}
\usepackage{wrapfig/wrapfig}
\usepackage[caption = false]{subfig}
\usetikzlibrary{decorations.markings,intersections,shapes,patterns}
\usepackage{hyperref}

\newif\iflong
\longtrue

\newif\ifshort
\shortfalse

\definecolor{overapproxreachablecolor}{RGB}{217, 210, 255}
\definecolor{overapproxboundaryreachablecolor}{RGB}{58, 26, 175}
\definecolor{reachablestatecolor}{RGB}{88, 204, 233}
\definecolor{badstatecolor}{RGB}{255, 88, 88}
\newcommand{\StateInDomain}[2]{\{#1\}}

\tikzset{%
  state/.style={draw=gray, fill=gray!10},
  reachable/.style={fill=reachablestatecolor},
  bad/.style={fill=badstatecolor},
  trans/.style={->, thick, draw=gray},
  convex/.style={very thick, draw=overapproxboundaryreachablecolor}
}

\newcommand*{\TikzGrid}[6]{%
  \draw[gray!50, thin, step=1] (#1, #3) grid (#2, #4);
  \draw[thick,->] (#1,0) -- (#2,0) node[right] {#5};
  \draw[thick,->] (0,#3) -- (0,#4) node[above] {#6};
}
\newcommand*{\TikzPosState}[2]{%
  \filldraw[state, reachable] (#1,#2) circle (0.2);
}
\newcommand*{\TikzPosStateConvex}[2]{%
  \filldraw[state, reachable, convex] (#1,#2) circle (0.2);
}
\newcommand*{\TikzNegState}[2]{%
  \filldraw[state, bad] (#1-0.2,#2-0.2) rectangle ++(0.4, 0.4);
}
\newcommand*{\TikzUnknownState}[2]{%
  \filldraw[state] (#1,#2) circle (0.2);
}
\newcommand*{\TikzUnknownStateConvex}[4]{%
  \filldraw[state,] (#1,#2) circle (0.2);
  \draw[draw=overapproxboundaryreachablecolor, line width=0.4mm] (#1,#2)+(-0.2:0.2) arc[start angle=#3, end angle=#4, radius=0.2];
}

\title{Data-driven Numerical Invariant Synthesis with Automatic Generation of Attributes\thanks{This work was supported in part by the french ANR project AdeCoDS.}}
\titlerunning{Data-driven Numerical Invariant Synthesis}
\author{Ahmed Bouajjani\inst{1} \and
  Wael-Amine Boutglay\inst{1,2}
  \and Peter Habermehl\inst{1}}

\authorrunning{A. Bouajjani et al.}

\institute{Universit\'e Paris Cit\'e, IRIF, Paris, France\\ \email{\{abou,boutglay,haberm\}@irif.fr} \and
Mohammed VI Polytechnic University, Ben Guerir, Morocco}

\begin{document}

\maketitle              

\begin{abstract}
We propose a data-driven algorithm for numerical invariant synthesis and verification. The algorithm is based on the ICE-DT schema for learning decision trees from samples of positive and negative states and implications corresponding to program transitions. The main issue we address is the discovery of relevant attributes to be used in the learning process of numerical invariants. We define a method for solving this problem guided by the data sample. It is based on the construction of a separator that covers positive states and excludes negative ones, consistent with the implications. The separator is constructed using an abstract domain representation of convex sets. The generalization mechanism of the decision tree learning from the  constraints of the separator allows the inference of general invariants, accurate enough for proving the targeted property. We implemented our algorithm and showed its efficiency.

\end{abstract}

\section{Introduction}

Invariant synthesis for program safety verification is a highly challenging problem. Many approaches exist for tackling this problem, including abstract interpretation, CEGAR-based symbolic reachability, property-directed reachability (PDR), etc. \cite{abstract-interpretation-1977,houdini-2001,cegar-2003,constraint-solving-2003,ic3-2011,pdr-2011,gpdr-2012,spacer-2014}. While those approaches are applicable to large classes of programs, they may have scalability limitations and fail to infer certain types of invariants, such as disjunctive invariants. Emerging data-driven approaches, following the active learning paradigm with various machine learning techniques, have shown their ability to solve efficiently complex instances of the invariant synthesis problem
\cite{ice-2014,ice-dt-2016,hornice-2018,svm-2012,svm-2017,code2inv-2020,loopinvgen-pie-2016}.
%
These approaches are based on the iterative interaction between a {\em learner} inferring candidate invariants from a {\em data sample}, i.e., a set of data classified either as positive examples, known to be reachable from the initial states and that therefore must be included in any solution, or negative examples, known to be predecessors of states violating the safety property and that therefore cannot be included in any solution, and a {\em teacher} checking the validity of the proposed solutions and providing counterexamples as feedback in case of non-validity.
One such data-driven approach is ICE \cite{ice-2014} which has shown promising results with its instantiation ICE-DT \cite{ice-dt-2016} that uses decision trees for the learning component. ICE is a learning approach tailored for invariant synthesis, where the feedback provided by the teacher can be, in addition to positive and negative examples, implications of the form $p \rightarrow q$ expressing the fact that if $p$ is in a solution, then necessarily $q$ should also be
included in the solution since there is a transition in the program from $p$ to $q$. 

The strength of data-driven approaches is the generalization mechanisms of their learning components, allowing them to find relevant abstractions from a number of examples without exploring the whole state space of the program. In the case of ICE-DT, this is done by a sophisticated construction of decision trees classifying correctly the known positive and negative examples at some point, and taking into account the information provided by the implications. These decision trees, where the tested attributes are predicates on the variables of the program, are interpreted as formulas corresponding to candidate invariants. 

However, to apply data-driven methods such as ICE-DT, one needs to have a pool of attributes that are potentially relevant for the construction of the invariant. This is actually a crucial issue. 
In ICE-DT, as well as in most data-driven methods, finding the predicates involved in the invariant construction is based on systematic enumeration of formulas according to some pre-defined templates or grammars. For instance, in the case of numerical programs, the considered patterns are some special types of linear constraints, and candidate attributes are generated by enumerating all possible values for the coefficients under some fixed bound. While such a brute-force enumeration can be effective in many cases, it represents, in general, an obstacle for both scalability and finding sufficiently accurate inductive invariants in complex cases. 

In this paper, we provide an algorithmic method for efficient generation of attributes for data-driven invariant synthesis for numerical programs manipulating integer variables. While enumerative approaches are purely syntactic and do not take into account the data sample, our method is guided by it. We show that this method, when integrated in the ICE-DT schema, leads to a new invariant synthesis algorithm outperforming state-of-the-art methods and tools. 

Our method for attributes discovery is based on, given an ICE data sample, computing a {\em separator} of it as a union of convex sets i.e., (1) it covers all the positive examples, (2) it does not contain any negative example, and (3) it is consistent with the implications (for every $p \rightarrow q$ in the sample, if the separator contains $p$, then it should also contain $q$). Then, the set of attributes generated is the set of all constraints defining the separator. However, as for a given sample there might be several possible separators, a question is which separators to consider. Our approach is guided by two requirements: (1) we need to avoid big pools of attributes in order to reduce the complexity of the invariant construction process, and (2) we need to avoid having in the pool constraints that are (visibly) unnecessary, e.g. separating positive examples in a region without any negative ones.
Therefore, we consider separators that satisfy the property that, whenever they contain two convex sets, it is impossible to take their convex union (smallest convex set containing the union) without including a negative example. 
%

To represent and manipulate algorithmically convex sets, we consider abstract domains, e.g., intervals, octagons, and polyhedra, as they are defined in the abstract interpretation framework and implemented in tools such as APRON \cite{apron-2009}. These domains correspond to particular classes of convex sets, defined by specific types of linear constraints. In these domains, the union operation is naturally over-approximated by the {\em join} operation that computes the best over-approximation of the union in the considered class of convex sets. Then, constructing separators as explained above can be done by iterative application of the join operation while it does not include negative examples.

Then, this method for generating candidate attributes can be integrated into the ICE-DT schema: in each iteration of ICE loop, given a sample, the learner (1) generates a set of candidate attributes from a separator of the sample, (2) builds a decision tree from these attributes and proposes it as a candidate invariant to the teacher. Then, the teacher (1) checks that the proposed solution is an inductive invariant, and if it is not (2) provides a counterexample to the learner, extending the sample that will be used in the next iteration.

Here a question might be asked: why do we need to construct a decision tree from the constraints of the separator and do not propose directly the formula defining the separator as a candidate invariant to the teacher. The answer is that the decision tree construction is crucial for generalization. Indeed, given a sample, the constructed separator might be too specialized to that sample and does not provide a useful inductive invariant (except for some simple cases). For instance, the constructed separator is a union of {\em bounded} convex sets (polytopes), while invariants are very often unbounded convex sets (polyhedra). The effect of using decision trees, in this case, is to select the relevant constraints and discard the unnecessary bounds, leading very quickly to an unbounded solution that is general enough to be an inductive invariant. Without this generalization mechanisms, the ICE loop will not terminate in such (quite common) cases. 

The integration of our method can be made tighter and more efficient by making the process of building separators incremental along the ICE iterations: at each step, after the extension of the sample by the teacher, instead of constructing a separator of the new sample from scratch, the parts of previously computed separators not affected by the last extension of the sample are reused.
%
%
%

We have implemented our algorithm and carried out experiments on the SyGuS-Comp'19 benchmarks. Our method solves significantly more cases than the tools LoopInvGen \cite{loopinvgen-pie-2016,loopinvgen-overfitting-2019}, CVC4 \cite{cvc4-2011,cvc4-sy-2019}, and Spacer \cite{spacer-2014}, as well as our implementation of the original ICE-DT \cite{ice-dt-2016} algorithm (with template-based enumeration of attributes), with very competitive time performances. 

\noindent {\bf Related work.}
Many learning-based approaches for the verification of numerical programs have
been developed recently. One of the earliest approaches is Daikon
\cite{daikon-2007}. Given a pool of formulas,
it computes likely invariants from program executions.
Later approaches were developed for the synthesis of sound invariants,
for example \cite{svm-2012} iteratively generates a set of reachable and bad states and classifies them with a combination of half-spaces computed using SVM.
In \cite{goem-concepts-2013}, the problem is reformulated as learning
geometric concepts in machine learning. 
The first instantiation of the ICE framework was based on a constraint solver \cite{ice-2014}. Later on, it was instantiated using the decision trees
learning algorithm \cite{ice-dt-2016}.
Both those instantiations require a fixed template
for the invariants or the formulas appearing in them.
LoopInvGen enumerates predicates on-demand using the approach introduced
in \cite{loopinvgen-pie-2016}.
This is extended to a mechanism with hybrid enumeration of several domains or grammars \cite{loopinvgen-overfitting-2019}.
Continuous logic networks were also used to tackle the problem in CLN2INV \cite{cln2inv-2020}. Code2Inv \cite{code2inv-2020}, the first approach to introduce general deep learning methods to program verification, uses a graph neural network to capture the program structure and reinforcement learning to guide the search heuristic of a particular domain.

The learning approach of ICE and ICE-DT has been generalized to solve problems given as constrained horn clauses (CHC) in Horn-ICE \cite{hornice-2018}
and HoICE \cite{hoice-2018}.
Outside the ICE framework, \cite{dd-chc-2018} proposed a learning approach for solving CHC using decision trees and SVM for the synthesis of candidate predicates from a set of reachable and bad states of the program. The limitation of the non-ICE-based approach is that when the invariant is not inductive, the
program has to be rerun, forward and backward, to generate more reachable and bad states.

In more theoretical work, an abstract learning framework for synthesis, introduced in \cite{alfs-2016}, incorporates the principle of CEGIS (counterexample-guided inductive synthesis). A study of overfitting in invariant synthesis was conducted in \cite{loopinvgen-overfitting-2019}. ICE was compared with IC3/PDR in terms of complexity in \cite{complexity-2020}. A generalization of ICE with relative inductiveness \cite{rice-2017} can implement IC3/PDR following the paradigm of active learning with a learner and a teacher.
        
Automatic invariant synthesis and verification has been addressed by many other techniques based on exploring and computing various types of abstract representations of reachable states (e.g., \cite{abstract-interpretation-1977,houdini-2001,cegar-2003,constraint-solving-2003,ic3-2011,pdr-2011,gpdr-2012,spacer-2014}).
Notice that, although we use abstract domains for representation and manipulation of convex sets, our strategy for exploring the set of potential invariants is different from the ones used typically in abstract interpretation analysis algorithms \cite{abstract-interpretation-1977}.


\section{Safety verification using learning of invariants}

This section presents the approach we use for solving the safety verification
problem. It is built upon the ICE framework \cite{ice-2014} and in
particular its instantiation with the learning of decision trees \cite{ice-dt-2016}. We first define the verification problem.

\subsection{Linear Constraints and Safety Verification}
\label{sec:form}

Let $X$ be a set of variables.
Linear formulas over $X$ are boolean combinations of linear constraints of the form $\sum_{i = 1}^n a_i x_i \leq b$
where the $x_i$'s are variables in $X$, the $a_i$'s are integer constants, and $b \in \mathbb{Z} \cup \{+\infty\}$. 
We use linear formulas to reason symbolically about programs with integer variables. 
Assume we have a program with a set of variables $V$ and let $n = |V|$. A state of the program is a vector of integers in $\mathbb{Z}^n$. Primed versions of these variables are used to encode the transition relation $T$ of the program: for each $v \in V$, we consider a variable $v'$ to represent the value of $v$ after the transition. Let $V'$ be the set of primed variables, and consider linear formulas over $V \cup V'$ to define the relation $T$. 

The {\em safety verification problem} consists in, given a set of safe states $\mathit{Good}$, deciding whether, starting from a set of initial states $\mathit{Init}$, all the reachable states by iterative application of $T$ are in $\mathit{Good}$. Dually, this is equivalent to decide if starting from $\mathit{Init}$, it is possible to reach a state in $\mathit{Bad}$ which is the set of unsafe states (the complement of $\mathit{Good}$). Assuming that the sets $\mathit{Init}$ and $\mathit{Good}$ can be defined using linear formulas, the safety verification problem amounts to find an adequate {\em inductive invariant} $I$, such that the three following formulas are valid:
\begin{eqnarray}
\mathit{Init}(V) & \; \Rightarrow \; & I (V) \label{inv:cond1} \\ 
\mathit{I} (V) & \; \Rightarrow \; & \mathit{Good} (V) \label{inv:cond2}\\
I (V) \wedge T (V, V') & \; \Rightarrow \; & I (V')\label{inv:cond3}
\end{eqnarray}

We are looking for inductive invariants
which can be expressed as a linear formula. In that case, the validity of the
three formulas
is decidable and can be checked with a standard SMT solver.

\subsection{The ICE learning framework}

ICE \cite{ice-2014}
follows the active learning paradigm to learn adequate inductive invariants
of a given program
and a given safety property.
It consists of an iteratively communicating \emph{learner} and a
\emph{teacher} (see Algorithm~\ref{algo:ice}).

      \begin{algorithm}[h]
           \SetKwInOut{Input}{Input}\SetKwInOut{Output}{Output}
            \Input{A transition system and a property: $(Init, T, Good)$}
            \Output{An adequate invariant or error}
            initialize ICE-sample $S=(S^+,S^-,S^\rightarrow)$\;
            \While{true}{
              $J  \gets \textsc{Learn}(S)$\;
              $(success,counterexample) \gets \textsc{is\_inductive}(J)$\;
                \lIf{$success$}{
                    \KwRet{$J$}
                }\Else{$S \gets \textsc{update}(S,counterexample)$\;
                         \lIf{$contradictory(S)$}{\KwRet{$error$}}}}
    \caption{The main loop of {ICE}.}
    \label{algo:ice}
\end{algorithm}

In each iteration, in line 3, the \emph{learner}, which does not
know anything about the program, synthesizes a candidate invariant (as a formula over the program variables) 
from a {\em sample} $S$ (containing information about program states)
which is enriched during the learning process.
Contrary to other learning methods, the sample $S$ not only contains
a set of {\em positive} states $S^+$ which should satisfy the invariant, and
a set of {\em negative} states $S^-$ which should not
satisfy the invariant, but it contains also a set of {\em implications} $S^\rightarrow$
of the form $s \rightarrow s'$ meaning that if $s$ satisfies the invariant, then $s'$
should satisfy it as well (because there is a transition from $s$ to $s'$ in the transition relation of the program). 
Therefore, an ICE-sample $S$ is a triple $(S^+,S^-,S^\rightarrow)$, where
to account for the information contained in implications, it is
imposed additionally that
\begin{equation}\label{prop:icesample}
\forall s \rightarrow s' \in S^\rightarrow:
\mbox{ if }s \in S^+\mbox{, then }s' \in S^+\mbox{, and if }s' \in S^{-}\mbox{, then }s \in S^{-}
\end{equation}
The sample is initially empty (or containing some states whose status, positive or negative, is known).
It is assumed that a candidate invariant $J$ proposed by the learner is  {\em consistent} with the sample, i.e. states in $S^+$ satisfy
the invariant $J$, the states in $S^-$ falsify it, and for implications $s \rightarrow s' \in S^\rightarrow$ it
is not the case that $s$ satisfies $J$ but not $s'$.
Given a candidate invariant $J$ provided by the \emph{learner}
in line 3, the \emph{teacher} who knows 
the transition relation $T$, checks if $J$ is an inductive invariant in line 4;
if yes, the process stops, an invariant has been found; otherwise 
a counterexample is provided and
used in line 7 to update the sample for the next iteration.
The teacher checks the three conditions an inductive invariant must
satisfy (see Section~\ref{sec:form}).
If (\ref{inv:cond1}) is violated the counterexample is a state $s$
which should be in the invariant because it is in $\mathit{Init}$.
Therefore $s$ is added to $S^+$.
If (\ref{inv:cond2}) is violated the counterexample is a state $s$ which
should not be in the invariant because it is not in $\mathit{Good}$
and $s$ is added to $S^-$.
If (\ref{inv:cond3}) is violated the counterexample is an implication
$s \rightarrow s'$ where if $s$ is in the invariant, $s'$ should also
be in it. Therefore  $s \rightarrow s'$ is added to $S^\rightarrow$.
In all three cases, the sample is updated to satisfy property
\ref{prop:icesample}. If this leads to a contradictory sample,
i.e. $S^+ \cap S^- \not= \emptyset$, the program is incorrect
and an error is returned.
Notice that obviously, in general, the loop is not guaranteed to
terminate.

\subsection{ICE-DT: Invariant Learning using Decision Trees}
\label{sec:icedt}

In \cite{ice-dt-2016}, the ICE learning framework 
is instantiated with a learn method, 
which extends classical decision tree learning algorithms
with the handling of implications. 
In the context of invariant synthesis,
{\em decision trees} are used to classify points from
a universe, which is the set of program states.
They are binary trees whose inner nodes are
labeled by predicates from a set
of attributes
and whose leaves are either $+$ or $-$. 
Attributes are (atomic) formulas over the variables of the program. They can be seen as boolean functions that the decision tree learning algorithm will compose to construct a classifier of the given ICE sample. In our case of numerical programs manipulating integer variables, attributes are linear inequalities.
Then, a decision tree can be seen naturally as a quantifier-free formula over program variables.

\begin{algorithm}[h]
      \SetKwProg{Proc}{Proc}{}{}
      \SetKwInOut{Input}{Input}\SetKwInOut{Output}{Output}
      \SetKwInOut{Global}{Global}
      \Input{An ICE sample $S = (S^+, S^-, S^\rightarrow)$}
      \Output{A formula}
      \Global{$Attributes$ initialized with $InitialAttributes$}
      \Proc{$\textsc{Learn}(S)$}{
            $(success,S_{Attr}) \gets \textsc{sufficient}(Attributes,S)$\;
            \While{$\neg success$}{
              $Attributes \gets \textsc{generateAttributes}(Attributes,S)$\;
              $(success,S_{Attr}) \gets \textsc{sufficient}(Attributes,S)$\;}
            \KwRet{$tree\_to\_formula(\textsc{Construct-Tree}(S_{Attr},Attributes))$}
      }
      \caption{\label{algo:learnupdateatt}The ICE-DT learner $\textsc{Learn}(S)$ procedure.}
\end{algorithm}

The main idea of the ICE-DT learner
(see Algorithm~\ref{algo:learnupdateatt})
is as follows.
Initially, the learner fixes a set of attributes (possibly empty)
which is kept in a global
variable and updated in successive executions of $\textsc{Learn}(S)$.
In line 2, given a sample, the learner
checks whether the current set of attributes is sufficient
to produce a decision tree corresponding to a formula consistent with
the sample.
If the check is successful the sample $S$ is changed 
to $S_{Attr}$ taking into account information gathered during the check
(see below for the details of $\textsc{sufficient}(Attributes,S)$).
If the check fails new attributes are generated
with $\textsc{generateAttributes}(Attributes,S)$
until success.
Then, a decision tree is constructed in line 6 from
the sample $S_{Attr}$ by
$\textsc{Construct-Tree}(S_{Attr},Attributes)$ which
we present below
(Algorithm~\ref{algo:ice-dt}).
It is transformed into a formula and
returned as a potential invariant.
Notice that in the main ICE loop of Algorithm~\ref{algo:ice}
the teacher then checks if this invariant is inductive or not.
If not, the original sample $S$ is updated and
in the next iteration the learner checks if the attributes
are still sufficient for the updated sample.
If not, the learner generates new attributes 
and proceeds with constructing another decision tree and so on.

An important question is how to choose $InitialAttributes$
and how to generate new attributes when needed.
In \cite{ice-dt-2016}, the set $InitialAttributes$ is for example
the set of octagons over program variables
with absolute values of constants bounded by $c \in \mathbb{N}$.
If these attributes are not sufficient to classify the sample, then
new attributes are generated simply by increasing the bound $c$ by $1$.
We use a different method described in detail in Section \ref{sec:genapp}.
We now describe how a decision tree can be constructed from an
ICE sample and a set of attributes.

\noindent{\bf Decision tree learning algorithms.}
The well-known standard decision tree learning algorithms like ID3 \cite{machine-learning-book-1997}
take as an input a sample containing points marked as positive or
negative of some universe and a fixed set $Attributes$.
They construct a decision tree 
by choosing as the root an attribute,
splitting the sample in two (one with all points satisfying the attribute
and one with the other points) and recursively constructing trees for
the two subsamples. At each step the attribute maximizing
the information gain computed using the entropy
of subsamples is chosen.
Intuitively this means that at each step, the attribute which
separates the ``best'' positive and negative points is chosen.
In the context of verification, exact classification is needed,
and therefore, all points in a leaf must be classified in a way consistent
with the sample.

In \cite{ice-dt-2016} this idea is extended to handle also
implications which is essential for an ICE learner. 
The basic algorithm to construct a tree (given as Algorithm~\ref{algo:ice-dt}
below)
gets as input an ICE sample $S=(S^+,S^-,S^\rightarrow)$
and a set of $Attributes$ and produces a decision tree
\emph{consistent} with the sample, which means that each point in
$S^+$ (resp. $S^-$) is classified as positive (resp. negative) and
for each implication $(s,s') \in S^\rightarrow$ it is not the case
that $s$ is classified as positive and $s'$ as negative.
The initial sample $S$ is supposed to be consistent.

\let\oldnl\nl
\newcommand{\nonl}{\renewcommand{\nl}{\let\nl\oldnl}}

\begin{algorithm}[h]\ifshort\scriptsize\fi
  \SetKwProg{Proc}{Proc}{}{}
            \SetKwInOut{Input}{Input}\SetKwInOut{Output}{Output}
            \Input{An ICE sample $S = (S^+, S^-, S^\rightarrow)$ and a set of $Attributes$.}
            \Output{A tree}
            \Proc{$\textsc{Construct-Tree}(S,Attributes)$}{
              Set~$G$~(partial~mapping~of~end-points~of~impl.~to~$\{\textsc{Positive},\textsc{Negative}\}$)~to~empty~\;
              Let $Unclass$ be the set of all end-points of implications in $S^\rightarrow$\;
              Compute the implication closure of $G$ w.r.t. $S$\;
              \KwRet{$\textsc{DecisionTreeICE}(\langle S^+, S^-, Unclass \rangle, Attributes)$}\;
            }

        \setcounter{AlgoLine}{5}
          \Proc{$\textsc{DecisionTreeICE}(Examples = \langle Pos, Neg, Unclass \rangle, Attributes)$}{

            Move all points of $Unclass$ classified as $\textsc{Positive}$
            (resp. $\textsc{Negative}$) to $Pos$ (resp. $Neg$)\;
            \uIf{$Neg = \emptyset$} {
              Mark all points of $Unclass$ in G as $\textsc{Positive}$\;
              Compute the implication closure of $G$ w.r.t. $S$\;           
              \KwRet{Leaf($+$)}\;
            } \uElseIf{$Pos = \emptyset$} {
              Mark all points of $Unclass$ in G as $\textsc{Negative}$\;
              Compute the implication closure of $G$ w.r.t. $S$\;
              \KwRet{Leaf($-$)}\;
            } \Else {
              $a \gets \textsc{choose}(Attributes,Examples)$\;
        Divide $Examples$ into two: $Examples_{a}$ with all points
              satisfying $a$ and $Examples_{\neg a}$ the others\;
        $T_{left} \gets \textsc{DecisionTreeICE}(Examples_a, Attributes\setminus\{a\})$\;
              $T_{right} \gets \textsc{DecisionTreeICE}(Examples_{\neg a}, Attributes\setminus\{a\})$\;
    \KwRet{$Tree(a,T_{left},T_{right})$}\;
            }
            }
        \caption{The ICE-DT decision-tree learning procedures.}
        \label{algo:ice-dt}
      \end{algorithm}

The learner is similar to the classical decision tree learning algorithms.
However, it has to take care of implications.
To this end, the learner also considers the set of points
appearing as end-points in the implications but not in $S^+$ and $S^-$.
These points are considered in the beginning as unclassified, and
the learner will either mark them $\textsc{Positive}$ or
$\textsc{Negative}$ during the construction as follows:
if in the construction of the tree a subsample is reached
containing only positive (resp. negative) points and
unclassified points (lines 8 and 12 resp.),
{\em all} these points are classified as positive
(resp. negative). To make sure that implications are still consistent,
the {\em implication closure} with the newly classified points is computed
and stored in the global variable $G$, a (partial mapping)
of end-points in $S^{\rightarrow}$ to $\{\textsc{Positive},\textsc{Negative}\}$.
The implication closure of $G$ w.r.t. $S$ is defined as:
If $G(s) = \textsc{Positive}$ or $s \in S^+$ and $(s,s') \in S^{\rightarrow}$ then also
$G(s') = \textsc{Positive}$. If $G(s') = \textsc{Negative}$ or $s' \in S^-$
and $(s,s') \in S^{\rightarrow}$ then also $G(s) = \textsc{Negative}$.

The set $Attributes$ is such that  
a consistent decision tree will always be found, i.e. the set $Attributes$
in line 17 is never empty (see below).
An attribute in a node is chosen with 
$\textsc{choose}(Attributes,Examples)$
returning an attribute $a \in Attributes$ with the highest
$gain$ according to $Examples$.
We do not give the details of this function.
In \cite{ice-dt-2016} several gain functions
are defined extending the classical gain function based on entropy with
the treatment of implications. We use the one which
penalizes cutting implications (like ICE-DT-penalty).

\noindent{\bf Checking if the set of attributes is sufficient.}
Here we show how the function $\textsc{sufficient}(Attributes,S)$
of Algorithm~\ref{algo:learnupdateatt} is implemented in
\cite{ice-dt-2016}. Two states $s$ and $s'$
are considered equivalent (denoted by $\equiv_{Attributes}$), 
if they satisfy the same attributes of $Attributes$.
One has to make sure that two equivalent states are never
classified in different ways by the tree construction algorithm.
This is done by the following procedure:
For any two states $s$, $s'$ with $s \equiv_{Attributes} s'$
which appear in the sample (as positive or negative or
end-points of the implications) 
two implications $s \rightarrow s'$ and $s' \rightarrow s$ are added
to $S^{\rightarrow}$ of $S$.

Then, the implication closure of the sample is computed starting
from an empty mapping $G$ (all end-points are initially unclassified).
If during the computation of the
implication closure one end-point is classified as
both $\textsc{Positive}$ and $\textsc{Negative}$, then 
$\textsc{sufficient}(Attributes,S)$ returns $(false,S)$
else it returns $(true,S_{Attr})$ where $S_{Attr}$ is obtained from
$S=(S^+,S^-,S^\rightarrow)$ by adding to $S^+$ the end-points of implications
classified as $\textsc{Positive}$ and to $S^-$ the end-points classified
as  $\textsc{Negative}$.

In \cite{ice-dt-2016} it is shown that this guarantees in general
that a tree consistent with the sample will always be constructed
regardless of the order in which attributes are chosen.
We illustrate now the ICE-DT learner on a simple example.

\begin{example}\label{ex:icedt}
Let $S=(S^+,S^-,S^\rightarrow)$ be a sample (illustrated
in Fig.~\ref{fig:separator-for-S}) with two-dimensional
states (variables $x$ and $y$): $S^+ = \{(1,1),$ $(1,4),$  $(3,1),$ $(5,1),$ $(5,4),$ $(6,1), $ $(6,4)\}$,
$S^- = \{(4,1),$ $(4,2),$ $(4,3),$ $(4,4)\}$,
$S^\rightarrow = \{(2,2) \rightarrow (2,3), (0,2) \rightarrow (4,0)\}$.
We suppose that $Attributes = \{ x \geq 1, x \leq 3, y \geq 1, y \leq 4,
x \geq 5, x \leq 6 \}$ is given. In Section~\ref{sec:genapp}
we show how to obtain this set from the sample.
The learner first checks that the set $Attributes$ is sufficient to construct
a formula consistent with $S$.
 The check succeeds and 
 we have among others that $(2,2)$ and $(2,3)$ and the surrounding positive
 states on the left are all equivalent w.r.t. $\equiv_{Attributes}$.
    Therefore after adding implications (which we omit for clarity in the following) and the computation of the implication closure both $(2,2)$ and $(2,3)$ 
   are added to $S^+$.
   Then, the construction of the tree is started with $Examples$
   containing $9$ positive, $4$ negative and $2$ unclassified states.
   Depending on the gain function an attribute is chosen.
   Here, it is $x \geq 5$, since it
   separates all the positive states on the right from the rest and
   does not cut the implication.
   The set $Examples$ is split into the states satisfying $x \geq 5$ and those
   which don't~: $Examples_{x \geq 5}$ and $Examples_{x < 5}$.
   $Examples_{x \geq 5}$ contains only positive states
   $\{(5,1),(5,4),(6,1),(6,4)\}$
   and the branch is finished
   whereas $Examples_{x < 5}$ contains
   the remaining positive, negative and unclassified states
   and the construction continues.
   The attribute $x \leq 3$ is chosen and $Examples_{x < 5}$ split
   in two.
   $Examples_{x < 5 \wedge x \leq 3}$ contains the positive states
   $\{(1,1),(1,4),(3,1),(2,2),(2,3)\}$ and one unclassified state $(0,2)$.
   Therefore, the algorithm marks $(0,2)$ as positive
   and as there is an implication 
   $(0,2) \rightarrow (4,0)$, the state $(4,0)$ is marked positive as well
   and a leaf node is returned.
   The other branch $Examples_{x < 5 \wedge x > 3}$ now contains negative states
   $\{(4,1),(4,2),(4,3),(4,4)\}$ and a positive state $(4,0)$.
   Therefore another attribute is needed.
   Finally, the algorithm returns a tree corresponding to the
   formula $x \geq 5 \vee (x < 5 \wedge x \leq 3) \vee (x < 5 \wedge x > 3
   \wedge y < 1)$.
\end{example}
   

\section{Linear Formulas as Abstract Objects}

Algorithm~{\ref{algo:learnupdateatt}} requires a set of attributes as input.
In section~\ref{sec:genapp}, we show how to generate these attributes from
the sample. For that purpose,
we use numerical abstract domains to represent and manipulate algorithmically
sets of integer vectors representing program states.
We consider standard numerical domains defined in \cite{interval-domain-1976,polyhedra-domain-1978,octagon-domain-2006} and implemented in tools such as APRON \cite{apron-2009}: Intervals, Octagons, and Polyhedra.

Given a set of $n$ variables $X$ and a linear formula $\varphi$ over $X$,
let $[\![ \varphi ]\!] \subseteq {\mathbb Z}^n$ be
the set of all integer points satisfying the formula.
Now, a subset of ${\mathbb Z}^n$ is called
\begin{itemize}
\item an \emph{interval}, iff it is equal to
  $[\![ \varphi ]\!]$ where $\varphi$ is a conjunction
  of constraints of the form 
  $\alpha \leq x \leq \beta$, where $x \in X$, $\alpha \in \mathbb{Z} \cup \{-\infty\}$ and $\beta \in \mathbb{Z} \cup \{+\infty\}$.
\item an \emph{octagon}, iff it is equal to
   $[\![ \varphi ]\!]$ where $\varphi$ is a conjunction
  of constraints of the form 
  ${\pm} \; x \; \pm \, y \leq \alpha$ where $x, y \in X$ and $\alpha \in  \mathbb{Z} \cup \{+\infty\}$.
\item a \emph{polyhedra}, iff it is equal to
   $[\![ \varphi ]\!]$ where $\varphi$ is a conjunction
  of linear constraints of the form $\sum_{i = 1}^n a_i x_i \leq b$ where
  $X = \{x_1,\ldots,x_n\}$ and for every $i$, $a_i \in \mathbb{Z}$, and $b \in \mathbb{Z} \cup \{+\infty\}$.
\end{itemize}

Now, we can define several abstract domains as
complete lattices
$A^{type}_X = \langle D^{type}_X, \sqsubseteq, \sqcup, \sqcap, \bot, \top \rangle$,
where $type$ is either $int$, $oct$ or $poly$ and
$D^{int}_X$ is the set of intervals,
$D^{oct}_X$ is the set of octagons and
$D^{poly}_X$ the set of polyhedra.

The relation $\sqsubseteq$ is set inclusion.
The binary operation $\sqcup$ (resp. $\sqcap$) is the {\em join} (resp. {\em meet}) operation that defines the smallest (resp. greatest) element in $D_X$
that contains (resp. contained in) the union (resp. the intersection) of the two composed elements. Finally $\bot$ (resp. $\top$) corresponds to the empty set
(resp. ${\mathbb{Z}^n}$). 


We suppose that we have a function $\mathit{Form^{type}}(d)$ which
given an element $d \subseteq \mathbb{Z}^n$
of the lattice provides us a formula $\varphi$
of the corresponding type such that $[\![ \varphi ]\!] = d$.
There are many ways to describe the set $d$ with a formula $\varphi$.
Therefore the function $\mathit{Form^{type}}(d)$ depends on the particular
implementation of the abstract domains.
We furthermore define $\mathit{Constr^{type}}(d)$ to be the set of linear
constraints of $\mathit{Form^{type}}(d)$.

We drop the superscript $type$ from all preceding definitions,
when it is clear from the context or when we define notions for all types.

All singleton subsets of $\mathbb{Z}^n$ are elements of the lattices
and for example, if $p = (x = 1, y =2)$, then, for the domains of Intervals, Octagons, and Polyhedra as implemented in APRON we have:
$\mathit{Constr^{int}} (\{p\}) = \{x \leq 1, x \geq 1, y \leq 2, y \geq 2\}$,
$\mathit{Constr^{oct}} (\{p\}) =  \{
x \geq 1, x \leq 1, y - x \geq 1, x + y \geq 3,
y \geq 2, y \leq 2, x + y \leq 3, x - y \geq -1 \}$
and $\mathit{Constr^{poly}} (\{p\})  = \{x = 1, y = 2\}$.

%
Notice, that in APRON while equality constraints are used in the Polyhedra
domain, these constraints are not explicit in the Interval and
Octagon domains.

An important fact about the three domains mentioned above is that,
each element of the lattice is the intersection of a convex
subset of $\mathbb{Q}^n$ with $\mathbb{Z}^n$.
To be able to reason about integer points from
{\em nonconvex} sets, we will use in the next section sets of sets.


\section{Generating Attributes from Sample Separators}
\label{sec:genapp}

We define in this section algorithms for generating a set of attributes that can be used for constructing decision trees representing candidate invariants.
Given an ICE sample, these algorithms are based on constructing separators of
the two sets of positive and negative states that are consistent with the implications in the sample. These separators are sets of intervals, octagons or polyhedra. The set of all constraints that define these sets
are collected as a set of attributes.

\subsection{Abstract Sample Separators}

Let  $S=(S^+,S^-,S^\rightarrow)$ be an ICE sample, and let $A_X = \langle D_X, \sqsubseteq, \sqcup, \sqcap, \bot, \top \rangle$ be an abstract domain.
Intuitively, a separator has sets containing all positive
states, not containing any negative state and is consistent with implications.
Formally,
an $A_{X}$-separator of $S$ is a set $\mathbb{S} \in 2^{D_X}$ such that
$\forall p \in S^{+}. \; \exists d \in \mathbb{S}. \; p \in d$ and
$\forall p \in S^{-}. \; \forall d \in \mathbb{S}. \; p \not\in  d$ and
$\forall p \rightarrow q \in S^\rightarrow. \; \forall d \in \mathbb{S}. \; (p \in d \implies (\exists d' \in \mathbb{S}. \; q \in  d'))$.

Given a set of positive states $S^+$, we define the basic
separator $\mathbb{S}_{basic}$ as $\{\{p\} \; | \; p \in S^+ \}$
where each state is alone in its set.
Our method for generating attributes for the learning process is based on computing a special type of separators called {\em join-maximal}. An $A_X$-separator $\mathbb{S}$ is join-maximal if is not possible to take the join of two of its elements without including a negative state:
$\forall d_1, d_2 \in \mathbb{S}.\; d_1 \neq d_2 \implies (\exists n \in S^-.\; n \in d_1 \sqcup d_2)$.

\begin{figure}[t]
\begin{center}
  \begin{tabular}{ >{\centering\arraybackslash}m{9.2em} >{\centering\arraybackslash}m{9.2em} >{\centering\arraybackslash}m{9.1em} >{\centering\arraybackslash}m{9.2em} }
       \begin{tikzpicture}[scale=0.28,state/.style={draw=gray, fill=gray!10},
    reachable/.style={fill=reachablestatecolor},
    bad/.style={fill=badstatecolor},
    trans/.style={->, thick, draw=gray}]
        \draw[gray!50, thin, step=1] (-1, -1) grid (7, 5);
        \draw[thick,->] (-1,0) -- (7,0) node[right] {$x$};
        \draw[thick,->] (0,-1) -- (0,5) node[above] {$y$};
       \draw [thick,dashed] (3,-1) -- (3,5);
      \draw [thick,dashed] (5,-1) -- (5,5);
      \draw [thick,dashed] (-1,1) -- (7,1);
     \TikzPosState{1}{1}
       \TikzPosState{1}{4}
        \TikzPosState{3}{1}
        \TikzPosState{5}{1}
        \TikzPosState{5}{4}
        \TikzPosState{6}{1}
        \TikzPosState{6}{4}
        \TikzNegState{4}{1}
        \TikzNegState{4}{2}
        \TikzNegState{4}{3}
        \TikzNegState{4}{4}
        \TikzUnknownState{2}{3}
        \draw[trans] (2, 2) -- (2,2.8) node[above] {};
        \TikzUnknownState{2}{2}
        \TikzUnknownState{4}{0}
       \draw[trans] (0, 2) -- (3.84,0.1) node[above] {};
        \TikzUnknownState{0}{2}
    \end{tikzpicture} &
    \begin{tikzpicture}[scale=0.28]
        \TikzGrid{-1}{7}{-1}{5}{$x$}{$y$}
        \path[draw=overapproxboundaryreachablecolor, fill=overapproxreachablecolor, fill opacity=0.5, very thick] (1,1) -- (1, 4) -- (3, 4) -- (3,1) -- (1, 1);
        \path[draw=overapproxboundaryreachablecolor, fill=overapproxreachablecolor, fill opacity=0.5, very thick] (5,1) -- (5, 4) -- (6, 4) -- (6,1) -- (5, 1);
        \TikzPosState{1}{1}
        \TikzPosState{1}{4}
        \TikzPosState{3}{1}
        \TikzPosState{5}{1}
        \TikzPosState{5}{4}
        \TikzPosState{6}{1}
        \TikzPosState{6}{4}
        \TikzNegState{4}{1}
        \TikzNegState{4}{2}
        \TikzNegState{4}{3}
        \TikzNegState{4}{4}
        \TikzUnknownState{2}{3}
        \draw[trans] (2, 2) -- (2,2.8) node[above] {};
        \TikzUnknownState{2}{2}
        \TikzUnknownState{4}{0}
        \draw[trans] (0, 2) -- (3.84,0.1) node[above] {};
        \TikzUnknownState{0}{2}
    \end{tikzpicture} & \begin{tikzpicture}[scale=0.28]
        \TikzGrid{-1}{7}{-1}{5}{$x$}{$y$}
        \path[draw=overapproxboundaryreachablecolor, fill=overapproxreachablecolor, fill opacity=0.5, very thick] (1,1) -- (1, 4) -- (3, 2) -- (3,1) -- (1, 1);
        \path[draw=overapproxboundaryreachablecolor, fill=overapproxreachablecolor, fill opacity=0.5, very thick] (5,1) -- (5, 4) -- (6, 4) -- (6,1) -- (5, 1);
        \TikzPosState{1}{1}
        \TikzPosState{1}{4}
        \TikzPosState{3}{1}
        \TikzPosState{5}{1}
        \TikzPosState{5}{4}
        \TikzPosState{6}{1}
        \TikzPosState{6}{4}
        \TikzNegState{4}{1}
        \TikzNegState{4}{2}
        \TikzNegState{4}{3}
        \TikzNegState{4}{4}
        \TikzUnknownState{2}{3}
        \draw[trans] (2, 2) -- (2,2.8) node[above] {};
        \TikzUnknownState{2}{2}
        \TikzUnknownState{4}{0}
        \draw[trans] (0, 2) -- (3.84,0.1) node[above] {};
        \TikzUnknownState{0}{2}
    \end{tikzpicture} & \begin{tikzpicture}[scale=0.28]
        \TikzGrid{-1}{7}{-1}{5}{$x$}{$y$}
        \path[draw=overapproxboundaryreachablecolor, fill=overapproxreachablecolor, fill opacity=0.5, very thick] (1,1) -- (1, 4) -- (2, 3) -- (3,1) -- (1, 1);
        \path[draw=overapproxboundaryreachablecolor, fill=overapproxreachablecolor, fill opacity=0.5, very thick] (5,1) -- (5, 4) -- (6, 4) -- (6,1) -- (5, 1);
        \TikzPosState{1}{1}
        \TikzPosState{1}{4}
        \TikzPosState{3}{1}
        \TikzPosState{5}{1}
        \TikzPosState{5}{4}
        \TikzPosState{6}{1}
        \TikzPosState{6}{4}
        \TikzNegState{4}{1}
        \TikzNegState{4}{2}
        \TikzNegState{4}{3}
        \TikzNegState{4}{4}
        \TikzUnknownState{2}{3}
        \draw[trans] (2, 2) -- (2,2.8) node[above] {};
        \TikzUnknownState{2}{2}
        \TikzUnknownState{4}{0}
        \draw[trans] (0, 2) -- (3.84,0.1) node[above] {};
        \TikzUnknownState{0}{2}
    \end{tikzpicture} \\
    (a) An ICE sample & (b) Intervals ($\texttt{int}$) & (c) Octagons ($\texttt{oct}$) & (d) Polyhedra ($\texttt{poly}$) \\
\end{tabular}
\end{center}
\caption{An ICE sample and its separators using different abstract domains.}
\label{fig:separator-for-S}
\end{figure}
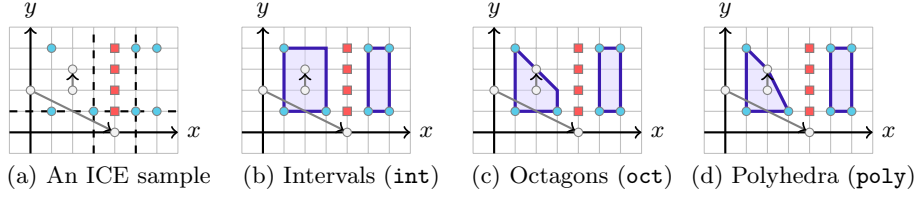

\begin{example}\label{ex:sep}
  Let us consider again the ICE sample $S$ given in Example~\ref{ex:icedt}.
Fig. \ref{fig:separator-for-S} shows the borders of join-maximal $A_X$-separators for $S$ for different abstract domains (Intervals $\texttt{int}$, Octagons $\texttt{oct}$, and Polyhedra $\texttt{poly}$).
\end{example}

\begin{remark}
  An ICE sample may have multiple join-maximal separators as Fig.~\ref{fig:multiple-sep} shows for the polyhedra domain.
  The method presented in the next section computes one of them non-deterministically.
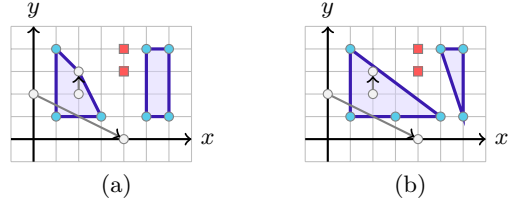
\begin{figure}[ht]
  \begin{center}
\begin{tabular}{ >{\centering\arraybackslash}m{1.5in} >{\centering\arraybackslash}m{1.5in}} 
    \begin{tikzpicture}[scale=0.3]
        \TikzGrid{-1}{7}{-1}{5}{$x$}{$y$}
        \path[draw=overapproxboundaryreachablecolor, fill=overapproxreachablecolor, fill opacity=0.5, very thick] (1,1) -- (1, 4) -- (2, 3) -- (3,1) -- (1, 1);
        \path[draw=overapproxboundaryreachablecolor, fill=overapproxreachablecolor, fill opacity=0.5, very thick] (5,1) -- (5, 4) -- (6, 4) -- (6,1) -- (5, 1);
        \TikzPosState{1}{1}
        \TikzPosState{1}{4}
        \TikzPosState{3}{1}
        \TikzPosState{5}{1}
        \TikzPosState{5}{4}
        \TikzPosState{6}{1}
        \TikzPosState{6}{4}
        \TikzNegState{4}{3}
        \TikzNegState{4}{4}
        \TikzUnknownState{2}{3}
        \draw[trans] (2, 2) -- (2,2.8) node[above] {};
        \TikzUnknownState{2}{2}
        \TikzUnknownState{4}{0}
        \draw[trans] (0, 2) -- (3.84,0.1) node[above] {};
        \TikzUnknownState{0}{2}
    \end{tikzpicture} & \begin{tikzpicture}[scale=0.3]
        \TikzGrid{-1}{7}{-1}{5}{$x$}{$y$}
        \path[draw=overapproxboundaryreachablecolor, fill=overapproxreachablecolor, fill opacity=0.5, very thick] (1,1) -- (1, 4) -- (5, 1) -- (1, 1);
        \path[draw=overapproxboundaryreachablecolor, fill=overapproxreachablecolor, fill opacity=0.5, very thick] (5, 4) -- (6, 4) -- (6,1) -- (5, 4);
        \TikzPosState{1}{1}
        \TikzPosState{1}{4}
        \TikzPosState{3}{1}
        \TikzPosState{5}{1}
        \TikzPosState{5}{4}
        \TikzPosState{6}{1}
        \TikzPosState{6}{4}
        \TikzNegState{4}{3}
        \TikzNegState{4}{4}
        \TikzUnknownState{2}{3}
        \draw[trans] (2, 2) -- (2,2.8) node[above] {};
        \TikzUnknownState{2}{2}
        \TikzUnknownState{4}{0}
        \draw[trans] (0, 2) -- (3.84,0.1) node[above] {};
        \TikzUnknownState{0}{2}
    \end{tikzpicture} \\
    (a) & (b) \\
\end{tabular}
\end{center}
\caption{Different join-maximal separators for a same sample.}
\label{fig:multiple-sep}
\end{figure}
\end{remark}

\subsection{Computing a Join-Maximal Abstract Separator}
\label{sec:algosimple}

We present in this section a basic algorithm for computing a join-maximal $A_X$-separator for a given sample $S$. 
Computing such a separator can be done iteratively starting from $\mathbb{S}_{basic}$, and at each step, choosing two elements $d_1$ and $d_2$  in the current separator such that $d_1 \sqcup d_2$ does not contain a negative state in $S^{-}$
(This can be checked using the meet operation $\sqcap$), and replacing $d_1$ and $d_2$ by $d_1 \sqcup d_2$. Then, if any element of the separator contains the source $p$ of an implication $p \rightarrow q$, which means that $p$ is considered now as a positive state, then since $q$ must also be considered as positive, the element $\{q\}$ must be added to the separator if $q$ is not already in some element of the current separator. When no new join operations (without including negative states) can be done, the obtained set is necessarily a join-maximal $A_X$-separator of $S$. This procedure corresponds to Algorithm \ref{algo:simple-algo}.

        \begin{algorithm}[h]\ifshort\scriptsize\fi
            \SetKwInOut{Input}{Input}\SetKwInOut{Output}{Output}\SetKwProg{Proc}{Proc}{}{}
            \Input{An ICE sample $S=(S^+,S^-,S^\rightarrow)$ and an abstract domain $A_X = \langle D_X, \sqsubseteq, \sqcup, \sqcap, \bot, \top \rangle$.}
            \Output{$\mathbb{S}$ a join-maximal $A_X$-separator of $S$.}
            \Proc{$\textsc{constructSeparator}(S, A_X)$}{
                $\mathbb{S} \gets \mathbb{S}_{basic}$ (* $= \{\{s\} \, \mid s \in S^{+} \}$ *) \;
                \While{$true$}{
                    \If{$\exists a,b \in \mathbb{S}.\, a \neq b \land \forall n \in S^{-}.\, n \notin a \sqcup b$}{
                        $\mathbb{S} \gets (\mathbb{S} \setminus \{a, b\}) \cup \{ a \sqcup b \}$ \;
                        \While{$\exists p \rightarrow q \in S^\rightarrow.\, \exists d \in \mathbb{S}.\, p \in d \land \forall d' \in \mathbb{S}.\, q \notin d'$} {
                          $\mathbb{S} \gets \mathbb{S} \cup \{ \{q\} \}$ \;
                        }
                    } \lElse {\textbf{break}}
                }
            }
                 \caption{Computing a join-maximal $A_X$-separator.}
            \label{algo:simple-algo}
        \end{algorithm}

Notice that instead of starting with the basic separator $\mathbb{S}_{basic}$
defined as above one can start with any separator $\mathbb{S}_{init} \supseteq
\mathbb{S}_{basic}$
whose additional sets contain only states which are known to be positive
(for example the initial states).


\begin{example}
  Consider again the sample $S$ of Example~\ref{ex:sep}.
  We show how the separators of $S$ in Fig. \ref{fig:separator-for-S} are constructed using Algorithm \ref{algo:simple-algo}. 
The algorithm starts from the basic separator 
$\mathbb{S}_{basic}$ where every positive state in $S$ is alone
(Fig. \ref{fig:step-sep-poly}(a)). It picks two elements in that separator, e.g. $\{d_1\}$ and $\{d_2\}$. As their join does not include negative states, $\{d_1\}$ and $\{d_2\}$ are replaced by $j_1 = \{d_1\} \sqcup \{d_2\}$
to get a new separator 
(Fig.~\ref{fig:step-sep-poly}(b)). 
Then, depending on the considered domain, different separators are obtained.
For Intervals, the join of $j_1$ and $\{d_3\}$ leads to the separator in Fig.~\ref{fig:separator-for-S}(a).
Notice that both ends of the implication $(2,2)\rightarrow(2,3)$ are included
in $j_1 \sqcup \{d_3\}$.
In the case of Octagons, the join of $j_1$ and $\{d_3\}$ is the set on
the left of Fig. \ref{fig:separator-for-S}(b). Again, both ends of the
implication $(2,2)\rightarrow(2,3)$ are included in $j_1 \sqcup \{d_3\}$. 
In the case of Polyhedra, $j_2 = j_1 \sqcup \{d_3\}$ is  the triangle shown in
Fig.~\ref{fig:step-sep-poly}(c). Since $(2,2)$ is included in $j_2$ but not
$(2,3)$, the element $\{(2,3)\}$ is added to the separator, leading to
the separator represented in Fig.~\ref{fig:step-sep-poly}(c).
In the next iteration, $j_2$ is joined with $\{d_8\}$ leading to the separator
shown in Fig.~\ref{fig:step-sep-poly}(d). 
Finally, a similar iteration of join operations 
leads to the rectangle including the four points, and this leads to the join-maximal separator of Fig.~\ref{fig:separator-for-S}.
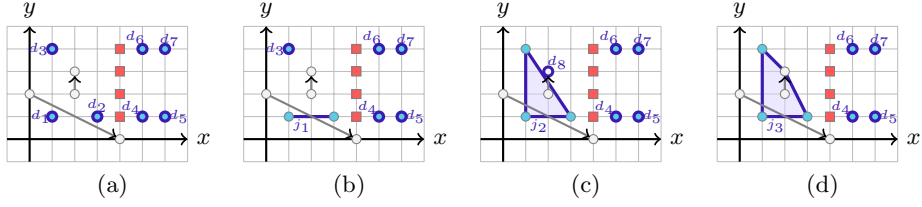
\begin{figure}[t]
\makebox[\textwidth][c]{
\begin{tabular}{ >{\centering\arraybackslash}m{1.2in} >{\centering\arraybackslash}m{1.2in} >{\centering\arraybackslash}m{1.2in} >{\centering\arraybackslash}m{1.2in} } 
    \begin{tikzpicture}[scale=0.3]
        \TikzGrid{-1}{7}{-1}{5}{$x$}{$y$}
        \node[text=overapproxboundaryreachablecolor] at (0.5,1) {\tiny $d_1$};
        \TikzPosStateConvex{1}{1}{0}{360}
        \node[text=overapproxboundaryreachablecolor] at (0.4,4) {\tiny $d_3$};
        \TikzPosStateConvex{1}{4}{0}{360}
        \node[text=overapproxboundaryreachablecolor] at (3,1.5) {\tiny $d_2$};
        \TikzPosStateConvex{3}{1}{0}{360}
        \node[text=overapproxboundaryreachablecolor] at (4.5,1.4) {\tiny $d_4$};
        \TikzPosStateConvex{5}{1}{0}{360}
        \node[text=overapproxboundaryreachablecolor] at (4.7,4.5) {\tiny $d_6$};
        \TikzPosStateConvex{5}{4}{0}{360}
        \node[text=overapproxboundaryreachablecolor] at (6.6,1) {\tiny $d_5$};
        \TikzPosStateConvex{6}{1}{0}{360}
        \node[text=overapproxboundaryreachablecolor] at (6.2,4.4) {\tiny $d_7$};
        \TikzPosStateConvex{6}{4}{0}{360}
        \TikzNegState{4}{1}
        \TikzNegState{4}{2}
        \TikzNegState{4}{3}
        \TikzNegState{4}{4}
        \TikzUnknownState{2}{3}
        \draw[trans] (2, 2) -- (2,2.8) node[above] {};
        \TikzUnknownState{2}{2}
        \TikzUnknownState{4}{0}
        \draw[trans] (0, 2) -- (3.84,0.1) node[above] {};
        \TikzUnknownState{0}{2}
    \end{tikzpicture} & \begin{tikzpicture}[scale=0.3]
        \TikzGrid{-1}{7}{-1}{5}{$x$}{$y$}
        \node[text=overapproxboundaryreachablecolor] at (1.6,0.6) {\tiny $j_1$};
        \path[draw=overapproxboundaryreachablecolor, fill=overapproxreachablecolor, fill opacity=0.5, very thick] (1,1) -- (3, 1);
        \TikzPosState{1}{1}
        \node[text=overapproxboundaryreachablecolor] at (0.4,4) {\tiny $d_3$};
        \TikzPosStateConvex{1}{4}{0}{360}
        \TikzPosState{3}{1}
        \node[text=overapproxboundaryreachablecolor] at (4.5,1.4) {\tiny $d_4$};
        \TikzPosStateConvex{5}{1}{0}{360}
        \node[text=overapproxboundaryreachablecolor] at (4.7,4.5) {\tiny $d_6$};
        \TikzPosStateConvex{5}{4}{0}{360}
        \node[text=overapproxboundaryreachablecolor] at (6.6,1) {\tiny $d_5$};
        \TikzPosStateConvex{6}{1}{0}{360}
        \node[text=overapproxboundaryreachablecolor] at (6.2,4.4) {\tiny $d_7$};
        \TikzPosStateConvex{6}{4}{0}{360}
        \TikzNegState{4}{1}
        \TikzNegState{4}{2}
        \TikzNegState{4}{3}
        \TikzNegState{4}{4}
        \TikzUnknownState{2}{3}
        \draw[trans] (2, 2) -- (2,2.8) node[above] {};
        \TikzUnknownState{2}{2}
        \TikzUnknownState{4}{0}
        \draw[trans] (0, 2) -- (3.84,0.1) node[above] {};
        \TikzUnknownState{0}{2}
    \end{tikzpicture} & \begin{tikzpicture}[scale=0.3]
        \TikzGrid{-1}{7}{-1}{5}{$x$}{$y$}
        \node[text=overapproxboundaryreachablecolor] at (1.6,0.6) {\tiny $j_2$};
        \path[draw=overapproxboundaryreachablecolor, fill=overapproxreachablecolor, fill opacity=0.5, very thick] (1,1) -- (3, 1) -- (1,4) -- (1,1);
        \TikzPosState{1}{1}
        \TikzPosState{1}{4}
        \TikzPosState{3}{1}
        \node[text=overapproxboundaryreachablecolor] at (4.5,1.4) {\tiny $d_4$};
        \TikzPosStateConvex{5}{1}{0}{360}
        \node[text=overapproxboundaryreachablecolor] at (4.7,4.5) {\tiny $d_6$};
        \TikzPosStateConvex{5}{4}{0}{360}
        \node[text=overapproxboundaryreachablecolor] at (6.6,1) {\tiny $d_5$};
        \TikzPosStateConvex{6}{1}{0}{360}
        \node[text=overapproxboundaryreachablecolor] at (6.2,4.4) {\tiny $d_7$};
        \TikzPosStateConvex{6}{4}{0}{360}
        \TikzNegState{4}{1}
        \TikzNegState{4}{2}
        \TikzNegState{4}{3}
        \TikzNegState{4}{4}
        \node[text=overapproxboundaryreachablecolor] at (2.4,3.4) {\tiny $d_8$};
        \TikzUnknownStateConvex{2}{3}{0}{360}
        \draw[trans] (2, 2) -- (2,2.8) node[above] {};
        \TikzUnknownState{2}{2}
        \TikzUnknownState{4}{0}
        \draw[trans] (0, 2) -- (3.84,0.1) node[above] {};
        \TikzUnknownState{0}{2}
    \end{tikzpicture} & \begin{tikzpicture}[scale=0.3]
        \TikzGrid{-1}{7}{-1}{5}{$x$}{$y$}
        \node[text=overapproxboundaryreachablecolor] at (1.6,0.6) {\tiny $j_3$};
        \path[draw=overapproxboundaryreachablecolor, fill=overapproxreachablecolor, fill opacity=0.5, very thick] (1,1) -- (1, 4) -- (2, 3) -- (3,1) -- (1, 1);
        \TikzPosState{1}{1}
        \TikzPosState{1}{4}
        \TikzPosState{3}{1}
        \node[text=overapproxboundaryreachablecolor] at (4.5,1.4) {\tiny $d_4$};
        \TikzPosStateConvex{5}{1}{0}{360}
        \node[text=overapproxboundaryreachablecolor] at (4.7,4.5) {\tiny $d_6$};
        \TikzPosStateConvex{5}{4}{0}{360}
        \node[text=overapproxboundaryreachablecolor] at (6.6,1) {\tiny $d_5$};
        \TikzPosStateConvex{6}{1}{0}{360}
        \node[text=overapproxboundaryreachablecolor] at (6.2,4.4) {\tiny $d_7$};
        \TikzPosStateConvex{6}{4}{0}{360}
        \TikzNegState{4}{1}
        \TikzNegState{4}{2}
        \TikzNegState{4}{3}
        \TikzNegState{4}{4}
        \TikzUnknownState{2}{3}{0}{360}
        \draw[trans] (2, 2) -- (2,2.8) node[above] {};
        \TikzUnknownState{2}{2}
        \TikzUnknownState{4}{0}
        \draw[trans] (0, 2) -- (3.84,0.1) node[above] {};
        \TikzUnknownState{0}{2}
    \end{tikzpicture} \\
    (a) & (b) & (c) & (d) \\
\end{tabular}
}
\caption{The first iterations of Algorithm \ref{algo:simple-algo} on the
  sample $S$ of Fig. \ref{fig:separator-for-S}}
\label{fig:step-sep-poly}
\end{figure}
\end{example}

\begin{remark}
In the best case Algorithm~\ref{algo:simple-algo}
  performs $|S^+|$ join and  $|S^+|(|S^-|+|S^{\rightarrow}|)$ meet operations (all pairs of points can be joined and all left end-points of implications
  are not in the new joined convex sets). In the worst case, it
  performs $O\big((|S^+|+|S^{\rightarrow}|)^2\big)$ join and
   $O\big((|S^+|+|S^{\rightarrow}|)^2(|S^-|+|S^{\rightarrow}|)\big)$ meet
  operations (at most $|S^-|+|S^{\rightarrow}|$ meets are needed
  to check if two sets can be joined and implications might add new points
  to $S^+$).
The cost of meet and join depends on the used abstract domain; it is polynomial for intervals and octagons, and exponential for polyhedra, in the number of variables.
Algorithm~\ref{algo:simple-algo} is not designed to compute
a join-max separator with a minimal number of convex sets as this would require
a potentially exponential number of meet and join operations. 
\end{remark}

\subsection{Integrating Separator Computation in ICE-DT}
\label{sec:integratesep}

We use the computation of a join-maximal separator
to provide an instance of~the function \textsc{generateAttributes} of ICE-DT in Algorithm~\ref{algo:learnupdateatt}.
Given a sample $S$, let $\mathbb{S}$ be the $A_X$-separator of $S$ computed by
$\textsc{constructSeparator}(S, A_X)$ defined by Algorithm \ref{algo:simple-algo}. We consider the set $\mathit{InitialAttributes}$ containing all the predicates that constitute the specification ($\mathit{Init}$ and $\mathit{Good}$) and those that appear in the programs (as tests in the conditional statements and while loops). 
Then, we define:
%
$\textsc{generateAttributes} ( S) = \mathit{InitialAttributes} \cup \bigcup_{d \in \mathbb{S}} \mathit{Constr}(d)$

\begin{remark}
  Several convex sets of the separator $\mathbb{S}$ might generate
  the same constraint and
the set of attributes generated in this way might contain
attributes which partition the state space in the same way
(e.g. $x\leq 0$ and $x\geq 1$, equivalent to $x>0$ over the integers).
We keep only one of them.
The number of attributes generated is at most linear in the number
of positive states in the sample $S$.
\end{remark}

\begin{figure}[ht]\begin{minipage}{\linewidth}\centering
            \begin{algorithm}[H]
              \textbf{int} j, k, t\;
              \textbf{assume}($j = 2 \land k = 0$)\;
              \While{$true$} {
                \leIf{$t=0$}{$j \gets j + 4$}{$j \gets j + 2; k \gets k + 1$}
              }
              \textbf{assert}($k = 0 \lor j = 2k + 2$)\;
            \end{algorithm}\end{minipage}
    \caption{Example program}
    \label{algo:ex-program}
\end{figure}

Notice that our function \textsc{generateAttributes}(S), contrary to the one used in the original ICE-DT (Algorithm \ref{algo:learnupdateatt}), does not expand a set of existing attributes, and therefore it only need the sample $S$ as argument. In fact, with our method for computing attributes, the ICE-DT schema can be simplified: the while loop in Algorithm \ref{algo:learnupdateatt} can be replaced by one single initial test on the condition of success. Indeed, each time the learner is called, it checks whether the set of attributes computed for the previous sample is sufficient to build a separator for the new sample. Only when it is not sufficient that the generation of a separator is performed. Then, the call of the {\sc sufficient} function afterward is needed to extend the sample
so that the construction of a decision tree can be done (see explanation in Section \ref{sec:icedt}), but it will necessarily succeed since in our case the set of attributes defines by construction a separator of the sample. 




\begin{example} \label{example:running-basic} 
Consider the program in Fig.~\ref{algo:ex-program}
whose set of variables is $X = \{j, k, t\}$. We use Polyhedra.
First, starting from an empty ICE-sample,
regardless of the attributes,
the learner proposes $\mathit{true}$ as an invariant
and $(5,1,0)$ is returned as a negative counterexample. Then, it
proposes $false$ and $(2,0,0)$ is returned as a positive counterexample.

Now, Algorithm~\ref{algo:simple-algo} is called to compute a separator for 
$S = (S^+ = \{(2,0,0)\},$ $S^- = \{(5,1,0)\}, S^\rightarrow=\emptyset)$.
Here, we use initially a separator $\mathbb{S}_{init}$
containing the set of states satisfying the initial condition
$j = 2\,\land\,k = 0$ denoted by $d_1$ in addition to $d_0$
where $d_0=\{(2,0,0)\}$.
Since $d_0 \subseteq d_1$, the algorithm returns the
join-maximal separator
$\mathbb{S} = \{d_1\}$ with $Constr^{poly}(d_1) = \{j = 2, k = 0\}$.

Using constraints from $\mathbb{S}$ as attributes, the learner constructs
the candidate invariant $k = 0$.
Then, the teacher provides an implication counterexample
$(0,0,1)\rightarrow(2,1,1)$. 
Now, without computing another separator (as the one it has is sufficient for the new sample), the learner proposes $j = 2 \land k = 0$ as an invariant, and
the implication counterexample $(2,0,1)\rightarrow(4,1,1)$ is returned (and since $(2,0,1)$ is an initial state, $(4,1,1)$ is also considered positive). 

Then, Algorithm \ref{algo:simple-algo} is called again to construct a separator for the sample $S = (S^+ = \{(2,0,0),(4,1,1)\}, S^- = \{(5,1,0)\}, S^\rightarrow=\{(0,0,1)\rightarrow(2,1,1), (2,0,1)\rightarrow(4,1,1)\})$.
Starting from a separator $\mathbb{S}_{init}=\{d_0,d_1,d_2\}$
with $d_2 = \{(4,1,1)\}$
it returns the join-maximal separator\\
\centerline{
  $\mathbb{S} = \{d_3\}\qquad Constr^{poly}(d_3) = \{2k + 2 = j, j \leq 4, j \geq 2\}$}

\noindent
Based on this separator, the learner proposes $2k + 2 = j$, $(2,0,0)\rightarrow(6,0,0)$ is given as a counterexample (and then, since $(2,0,0)$ is in $S^{+}$, $(6,0,0)$ is considered positive). Then, from 
$\mathbb{S}_{init} = \{d_0,d_1,d_2,d_4\}$
with $d_4 = \{(6,0,0)\}$
a new separator $\mathbb{S}$ is constructed 

\centerline{$\mathbb{S} = \{d_5\} \qquad Constr^{poly}(d_5) = \{j + 2k \leq 6, k \geq 0, j \geq 2k + 2\}$}
\noindent
leading to a new candidate invariant:   $j + 2k \leq 6 \land j \geq 2k + 2$. The teacher returns at this point the negative state $(0,-2,0)$. The attributes of $\mathbb{S}$ are still sufficient to construct a decision tree for the sample. Then, the learner proposes $j + 2k \leq 6 \land k \geq 0 \land j \geq 2k + 2$, and the teacher returns the counterexample $(3,0,1)\rightarrow(5,1,1)$ (and since $(5,1,1)$ is a negative state, $(3,0,1)$ is considered negative).  
The current sample $S$ is now 
$(S^+ = \{(2,0,0),(4,1,1),(6,0,0)\},$  $S^- = \{(5,1,0), (5,1,1), (3,0,1), (0,-2,0)\}$, $S^\rightarrow=\{(0,0,1) \rightarrow (2,1,1),$  $(2,0,1) \rightarrow (4,1,1),$
$(2,0,0) \rightarrow (6,0,0), (3,0,1) \rightarrow (5,1,1)\})$.

%
%
Then, from $ \mathbb{S}_{init} = \{d_0,d_1,d_2,d_4\}$, a join-maximal separator is constructed

\centerline{$\mathbb{S} = \{d_3, d_4\} \qquad Constr^{poly}(d_4) = \{j=6, t = 0, k = 0\}$}
\noindent Some iterations later, using only the attributes of the last $\mathbb{S}$,
the learner generates the inductive invariant
$(t = 0\,\land\, 2 \leq j\, \land \,k = 0)\, \lor\, (t \neq 0 \, \land \,2 \leq j\, \land \,2k + 2 = j)$
\end{example}

\subsection{Computing Separators Incrementally}

Algorithm \ref{algo:simple-algo} of Section~\ref{sec:algosimple} always starts from the initial separator, regardless of what has been done in the previous iterations of the ICE learning process. Here, we present an incremental approach to exploit the fact that adding a counterexample to the sample may modify the separator only locally allowing parts of separators computed in previous iterations to be reused.  
The basic idea is to store the history of the separator computation along the ICE iterations, and update it according to the new counterexamples discovered at each step. 

\vspace{5pt}\noindent{\bf {The Algorithm}}.
We use an abstract stack data structure to represent the history of separators.  Along the iterations of the ICE learning algorithm, an increasing sequence of samples $S_i$'s is considered (at each iteration it is enriched by the new counterexample provided by the teacher). Then, at each step $i$, a join-maximal separator $\mathbb{S}_i$ of the sample $S_i$ is computed and stored in the stack. Notice that at a given step $i$, separators of index $j < i$ are not necessarily separators of $S_i$ since they may not cover all positive points of $S_i$. Therefore, we introduce the following notion: a \textit{partial $A_X$-separator} of a sample $S$ is a set $\mathbb{S} \in 2^{D_X}$ such that $\forall p \in S^-.\, \forall d \in \mathbb{S}.\, p \notin d$.

Now, to compute the separator  $\mathbb{S}_i$, we start from one of the partial separators in the stack, the most recent one that is not affected by the last update of the sample. When the sample at step $i$ is extended with positive states, $\mathbb{S}_i$ can be computed directly from $\mathbb{S}_{i-1}$. However, when the sample is extended with negative states, this might require reconsidering several previous steps since some of the elements (convex sets) of their separators might contain states that are (discovered now to be) negative. In that case, we must return to the step of the greatest index $j < i$ (i.e., the last step before $i$) such that $\mathbb{S}_j$ is a partial separator of $S_i$ (i.e., the new knowledge about the negative states does not affect the computed separation at step $j$). By the fact that the sequence of samples is increasing, it is indeed correct to consider the biggest $j < i$ satisfying the property above. Therefore, the separator $\mathbb{S}_i$ is computed starting from $\mathbb{S}_j$ augmented with all the positive states in $S^+_i \setminus S^+_j$.

This leads to Algorithm \ref{algo:inc-abstract-algo}. We use in its description a stack $P$ 
supplied with the usual operations: $P.head()$ returns the top element of the stack, $P.\text{pop}()$ removes and returns the top element of the stack, and $P.\text{push}(e)$ inserts an element $e$ at the top position of the stack.
\ifshort
A refined version of Algorithm \ref{algo:inc-abstract-algo} is presented in
the full paper \cite{https://doi.org/10.48550/arxiv.2205.14943} where the backtracking phase is made more effective: We attach information to each join-created object in order to track its join-predecessors (objects involved in its creation) in the stack. 
\fi
\iflong
A refind version of Algorithm \ref{algo:inc-abstract-algo} is presented in
Section \ref{sec:graph-algo}  where the backtracking phase is made more effective: We attach information to each join-created object in order to track its join-predecessors (objects involved in its creation) in the stack. 
\fi

\newcommand{\IncFuncName}{constructSeparatorInc}

\begin{figure*}[ht]
\ifshort\scriptsize\fi
    \centering
    \begin{minipage}{\linewidth}
        \begin{algorithm}[H]
            \SetKwInOut{Input}{Input}\SetKwInOut{Output}{Output}\SetKwInOut{Requirement}{Requires}\SetKwInOut{Global}{Global}\SetKwProg{Proc}{Proc}{}{}
            \Global{$P = \{\emptyset\}$ a stack of partial separators.}
            \Proc{\textsc{\IncFuncName}($S_i=(S_i^+,S_i^-,S_i^\rightarrow), A_X$)}{
              \tcp{backtracking}
              \While{$true$} {
                \If{$\exists n \in S_i^-.\; \exists d \in P.head(). \; n \in d$} {
                  $P.\text{pop}()$\;
                } \lElse{\textbf{break}}
              }
              \tcp{expansion}
              $\mathbb{S} \gets P.head()$\;
              $add \gets \{p \in S_i^+ \, \mid \, \forall d \in \mathbb{S}.\, p \notin d\} \cup \{q \, \mid \, \exists p \rightarrow q \in S_i^\rightarrow. \, \exists d' \in \mathbb{S}.\, p \in d' \land \forall d'' \in \mathbb{S}.\, q \notin d''\}$\;
              \While{$\exists s \in add$} {
                $add \gets add \setminus \{s\}$\;
                \If{$\exists d \in \mathbb{S}.\, \forall n \in S_i^-.\, n \notin d \sqcup \StateInDomain{s}{A_X}$} {
                  \textbf{let} $o = d \sqcup \StateInDomain{s}{A_X}$\;
                  $\mathbb{S} \gets (\mathbb{S} \setminus \{d\}) \cup \{ o \}$\;
                  \For{$p \rightarrow q \in S_i^\rightarrow$ s.t. $p \in o \land \forall d' \in \mathbb{S}.\, q \notin d'$}{
                    $add \gets add \cup \{q\}$
                  }
                } \Else {
                  $\mathbb{S} \gets \mathbb{S} \cup \{ \StateInDomain{s}{A_X} \}$\;
                  \For{$p \rightarrow q \in S_i^\rightarrow$ s.t. $p = s \land \forall d' \in \mathbb{S}.\, q \notin d'$}{
                    $add \gets add \cup \{q\}$
                  }
                }
              }
              $P.\text{push}(\mathbb{S})$\;
              \KwRet{$\mathbb{S}$}\;
            }
            \vspace*{2ex}
            \caption{Incremental computation of an $A_X$-separator of a sample $S$.}
            \label{algo:inc-abstract-algo}
        \end{algorithm}
    \end{minipage}
\end{figure*}

\vspace{5pt}\noindent{\bf Integration to ICE-DT.}
The function $\textsc{\IncFuncName}$ can be integrated to the ICE-DT algorithm
just as the function $\textsc{constructSeparator}$ in Section \ref{sec:integratesep}, by using it to implement the function
$\sc{generateAttributes}$ of the learner. But this time, the learner is more efficient in computing the separator from which the attributes are extracted.

\begin{example} 
Consider again the program in Fig.~\ref{algo:ex-program} of Example~\ref{example:running-basic}.
The two first iterations are similar to the ones described in Example~\ref{example:running-basic}. Then, the obtained
sample is $S = (S^+ = \{(2,0,0)\}, S^- = \{(5,1,0)\}, S^\rightarrow=\emptyset)$. Starting from the empty separator, Algorithm~\ref{algo:inc-abstract-algo} computes the separator $\mathbb{S}_1 = \{d_1\}$ where
$Constr^{poly}(d_1) = \{j = 2, k = 0\}$.
%
%
Then, the learner proceeds as in the previous example to get the sample $S = (S^+ = \{(2,0,0),(4,1,1)\}, S^+ = \{(5,1,0)\}, S^\rightarrow=\{(0,0,1)\rightarrow(2,1,1), (2,0,1)\rightarrow(4,1,1)\})$. 
To build a separator of $S$, Algorithm~\ref{algo:inc-abstract-algo} starts from $\mathbb{S}_1$ and produces $\mathbb{S}_2 = \{d_3\}$ where $d_3 = d_1 \sqcup \{(4,1,1)\}$.

Similarly, when the counterexample $(2,0,0)\rightarrow(6,0,0)$ is obtained, the algorithm starts directly from $\mathbb{S}_2$ to produce $\mathbb{S}_3 = \{d_5\}$  where $d_5 = d_3 \sqcup \{(6,0,0)\}$.

After two more iterations, the sample is the same as $S'$ in
Example~\ref{example:running-basic}.
At this point, $\mathbb{S}_3$ cannot be used to construct a separator for $S$ since $d_5$ includes the negative state $(3,0,1)$. 
Then, the algorithm removes $\mathbb{S}_3$ from the stack. It checks that $\mathbb{S}_2$ is a partial separator of $S$, which is indeed the case. Then, it constructs a 
new separator $\mathbb{S}_4$ based on $\mathbb{S}_2$ by expanding it with the counterexamples received after the construction of $\mathbb{S}_2$ (the negative state $(0,-2,0)$ and the implications $(2,0,0)\rightarrow(6,0,0)$ and $(3,0,1)\rightarrow(5,1,1)$): $\mathbb{S}_4 = \{d_3, d_6\}$ where $Constr^{poly}(d_6) = \{t = 0, k = 0, j = 6\}$. The rest of the execution proceeds as with
Algorithm~\ref{algo:simple-algo}.
Here, the advantages of the incremental method are: (1) while positive examples are added the separators are simply expanded, and (2) when a negative example at step 4 is added, only one join operation has to be undone. 
\end{example}

\iflong
\subsection{A Refined Version of the Incremental Algorithm}
\label{sec:graph-algo}
Algorithm \ref{algo:inc-abstract-algo}  includes a backtracking phase where it searches for the most recent partial separator of $S$ in the stack. 
To make this backtracking phase more effective, we attach some information to the separators stored in the stack and to their elements. We associate with each newly created object an index that corresponds to the number of the current iteration. Assume the algorithm is at step $i$, and that a negative state $n$ is discovered such that $n \in d$ for some $d \in \mathbb{S}_{i-1}$. Then, knowing the index of $d$, say $j$ with $j < i-1$, allows to jump directly to $\mathbb{S}_{j-1}$ in the backtracking process. If the index of $d$ is $i-1$, tjhis means that $d = d_1 \sqcup d_2$ for some $d_1$ and $d_2$ in $\mathbb{S}_{i-2}$. Then, if one of these two objects contains $n$, then it is possible to jump back to step just before its creation using its index, and continue the backtracking process from there. So, in order to do this, we need to associate also with each object the set of its parents, i.e., objects that were joined for its creation, when they exist (since an object that be created initially as a singleton from a program state). 

In the following, a \textit{node} $N$ ($\in \mathcal{N}$, the set of nodes) is a structure with an abstract object ($object(N)$), an index ($index(N)$), and a set of nodes $parents(N)$ corresponding to its parents (or empty). Sets of nodes are called {\em forests}. Let $\mathcal{F} = 2^{\mathcal{N}}$ be the set of forests. We use forests to represent sets of nodes such that the set of their objects are separators.  For a forest $\mathbb{F} \in \mathcal{F}$, let 
$Separator(\mathbb{F}) = \{ \mathit{object}(N) \, \mid \, N \in \mathbb{F} \}$. 

The algorithm uses a stack of indexed forests: a pair of a forest with the index of the step at which it has been stored. This allows to express the backward jumps during the backtracking phase. 

\begin{figure*}[t]
    \centering
    \begin{minipage}{\linewidth}
        \begin{algorithm}[H]
            \SetKwInOut{Input}{Input}\SetKwInOut{Output}{Output}\SetKwInOut{Requirement}{Requires}\SetKwInOut{Global}{Global}\SetKwProg{Proc}{Proc}{}{}
            \Global{$P = \{(\emptyset, 0) \}$ a stack of indexed forests.}
            \Proc{\textsc{\IncFuncName}($S_i=(S_i^+,S_i^-,S_i^\Rightarrow, A_X)$)}{
              \tcp{backtracking}
              \While{$true$} {
                \If{$\exists n \in S_i^-.\; \exists N \in forest(P.head()). \; n \in object(N)$} {
                  $tmp \gets N$\;
                  \While{$n \in object(tmp)$} {
                    \If{$\exists N' \in parents(tmp).\; n \in object(N')$} {
                      $tmp \gets N'$
                    } \lElse {
                      \textbf{break}
                    }
                  }
                  \lWhile{$index(P.head()) \geq index(tmp)$} {
                    $P.\text{pop}()$
                  }
                } \lElse {\textbf{break}}
              }
              \tcp{expansion}
              $\mathbb{F} \gets forest(P.head())$\;
              $add \gets \{p \in S_i^+ \, \mid \, \forall N \in \mathbb{F}.\, p \notin object(N)\} \cup \{q \, \mid \, \exists p \rightarrow q \in S_i^\Rightarrow. \, \exists N' \in \mathbb{F}.\, p \in object(N') \land \forall N'' \in \mathbb{F}.\, q \notin object(N'')\}$\;
              \While{$\exists s \in add$} {
                $add \gets add \setminus \{s\}$\;
                \If{$\exists N \in \mathbb{F}$ s.t. $\forall n \in S_i^-.\, n \notin object(N) \sqcup \StateInDomain{s}{A_X}$} {
                  \textbf{let} $o = object(N) \sqcup \StateInDomain{s}{A_X}$\;
                  $\mathbb{F} \gets (\mathbb{F} \setminus \{N\}) \cup \{ (o, i, \{N, (\StateInDomain{s}{A_X}, i, \emptyset) \} ) \}$\;
                  \For{$p \rightarrow q \in S_i^\Rightarrow$ s.t. $p \in o \land \forall N' \in \mathbb{F}.\, q \notin object(N')$}{
                    $add \gets add \cup \{q\}$
                  }
                } \Else {
                  $\mathbb{F} \gets \mathbb{F} \cup \{ (\StateInDomain{s}{A_X}, i, \emptyset ) \}$\;
                  \For{$p \rightarrow q \in S_i^\Rightarrow$ s.t. $p = s \land \forall N' \in \mathbb{F}.\, q \notin object(N')$}{
                    $add \gets add \cup \{q\}$
                  }
                }
              }
              $P.\text{push}((\mathbb{F}, i))$\;
              \KwRet{$Separator(\mathbb{F})$}\;
            }
            \vspace*{2ex}
            \caption{A refinement of Algorithm \ref{algo:inc-abstract-algo}.}
            \label{algo:inc-algo-with-graph}
        \end{algorithm}
    \end{minipage}
\end{figure*}
\fi

\section{Experiments}

\newcommand{\ImplName}{\textsc{NIS}}

We have implemented the prototype tool \ImplName\
(Numerical Invariant Synthesizer) using our method for attribute synthesis with the ICE-DT schema. \ImplName\ written in C++
is configurable with an abstract domain for the manipulation of abstract objects. It uses Z3 \cite{z3-2008} for SMT queries and APRON's \cite{apron-2009} abstract domains.

\medskip

We compare our implementation with ICE-DT\footnote{The original ICE-DT tool \cite{ice-dt-2016} does not support programs in the SyGuS format. Here we use our own implementation of ICE-DT. It shares with \ImplName\ all the components (teacher, decision tree learning algorithm with implications) except that attribute discovery is enumerative.},  LoopInvGen, CVC4, and Spacer\footnote{Spacer does not support programs in the SyGuS format; a wrapper is written in C++ that converts a SyGuS program to a CHC problem and supplies it to Spacer via the Z3 FixedPoint API.}. LoopInvGen is a data-driven invariant inference tool based on a syntactic enumeration of candidate predicates \cite{loopinvgen-pie-2016,loopinvgen-overfitting-2019}. 
It is written in OCaml and uses Z3 as an SMT solver. 
CVC4 uses an enumerative refutation-based approach \cite{cvc4-2011,cvc4-sy-2019}. 
It is written in C++ and it includes an SMT solver. 
Spacer is a PDR-based CHC solver \cite{spacer-2014}, written in C++ and integrated in Z3.

\begin{figure}[t]

\begin{minipage}{.5\linewidth}
  \ifshort\scriptsize\fi
  \centering
\begin{tabular}{|| c || c | c || c ||}
\hline
\multirow{2}{*}{Tool} & \multicolumn{2}{c||}{Solved} & \multirow{2}{*}{Total} \\\cline{2-3}
     & safe & unsafe & \\
\hline
\hline
ICE-DT     & 111 & 11 & 122 \\
LoopInvGen & 130 & 8  & 138 \\
CVC4       & 129 & -  & 129 \\
Spacer     & 118 & 18 & 136 \\
\hline
\hline
\ImplName($\texttt{int}$)   & 106 & 17 & 123 \\
\ImplName($\texttt{oct}$)   & 122 & 14 & 136 \\
\ImplName($\texttt{poly}$)  & 137 & 17 & 154 \\
\hline
\hline
\ImplName(VB) & 143 & 17 & 160 \\
\hline
\end{tabular}
\end{minipage}
\begin{minipage}{.5\linewidth}
  \ifshort\scriptsize\fi
  \centering
\begin{tabular}{|| c || c | c || c ||}
\hline
                           & \ImplName($\texttt{int}$) & \ImplName($\texttt{oct}$) & \ImplName($\texttt{poly}$) \\\hline
\ImplName($\texttt{int}$)  &                         - &                         7 &                          2 \\\hline
\ImplName($\texttt{oct}$)  &                        13 &                         - &                          5 \\\hline
\ImplName($\texttt{poly}$) &                        31 &                        18 &                          - \\\hline
\end{tabular}
\end{minipage}

\caption{Benchmark results and comparison of \ImplName\ wrt. different abstract domains.}
\label{table:benchmark}\label{table:pairwise-comp}
\end{figure}

The evaluation was done on 164 linear integer arithmetic (LIA) programs\footnote{Other programs from SyGuS-Comp'19 have not been taken into account in our evaluations as they are boolean programs with integer variables for encoding nondeterminism or artificial programs augmented with useless variables and statements.}
from SyGuS-Comp'19. They have a number of variables ranging from 2 to 10. The experiments were carried out using a timeout of 1800s (30 minutes) for each example. They were conducted on a machine with 4 CPUs Intel(R) Xeon(R) 2,13GHz, 16 cores, and 128 Go RAM running Linux CentOS 7.9.


Figure~\ref{table:benchmark} shows the number of safe and unsafe solved programs by each tool. The instance of our approach using the Polyhedral abstract domain solves 154 programs out of 164, and the virtual best of our approach with the three abstract domains Intervals, Octagons, and Polyhedra, solves 160 programs out of 164. Two of the remaining examples require handling quantifiers, which cannot be done with the current implementation. The two others have not been solved with any of the four tools we considered.

  These results show that globally our approach is powerful and is able to solve a significant number of cases that are not solvable by other tools. Interestingly, using different abstract domains leads to incomparable performances: although with polyhedra more cases are solvable, there are some cases that are uniquely solvable with intervals or octagons. Also, while operations on intervals and octagons have a lower complexity than on polyhedra, this is compensated with the fact that polyhedra are more expressive. Indeed, their expressiveness allows in many cases to find quickly invariants for which a less expressive domain requires much more iterations to be learned. Figure~\ref{table:pairwise-comp} shows the number of programs that can be solved using a particular abstact domain but not with another. Polyhedra are globally superior, but the three domains are complementary.

Compared to the other tools, the bottleneck of ICE-DT and also of LoopInvGen is the number of predicates that are generated using enumeration. Our approach avoids the explosion of the size of the attribute pool by guiding their discovery with the data sample, and reducing the size (by replacing objects by their join) of the computed separators from which constraints are extracted. Concerning CVC4, it uses enumerative refutation techniques, which are also subject to an explosion problem. Moreover, CVC4 does not allow to solve the cases of unsafe programs. The performances of Spacer depend on the ability to generalize the set of predecessors computed using the model-based projection and the interpolants used for separation from bad states in the context of IC3/PDR. While this is done efficiently in general, there are cases where this process can lead to fastidious computations while our technique can be much faster using a small number of join operations of positive states.


The scatter plots shown in Fig. \ref{fig:runtime-plot} compare the execution times of our approach using Polyhedra abstract domain \ImplName($\texttt{poly}$) with LoopInvGen, CVC4 and Spacer. (A timeout of 1800s is used for each example.) They show that \ImplName($\texttt{poly}$) is in general faster than both LoopInvGen and CVC4, and that it has comparable performances in terms of execution time with Spacer. We have also compared the original ICE-DT, based on enumerative attribute generation using octagonal templates (as in \cite{ice-dt-2016}) with \ImplName($\texttt{oct}$). The comparison shows that our tool is significantly faster (see the bottom right subfigure of Fig. \ref{fig:runtime-plot}). 

\begin{figure}[t]
  \centering
  \captionsetup{justification=centering}
\makebox[0.8\linewidth][c]{%
\subfloat{
\scalebox{0.2}{\input{vsLoopInvGen.pgf}}
}\hfil
\subfloat{
\scalebox{0.2}{\input{vsCVC4.pgf}}
}
}
\makebox[0.8\linewidth][c]{
\subfloat{
\scalebox{0.2}{\input{vsSpacer.pgf}}
}\hfil
\subfloat{
\scalebox{0.2}{\input{vsICE-DT.pgf}}
}
}
\caption{Runtime of \ImplName($\texttt{poly}$) vs. LoopInvGen, CVC4, and Spacer,}{and \ImplName($\texttt{oct}$) vs. ICE-DT.}
\label{fig:runtime-plot}
\end{figure}
\section{Conclusion}

We have defined an efficient method for generating relevant predicates for the learning process of numerical invariants. The approach is guided by the data sample built during the process and is based on constructing a separator of the sample. The construction consists of an iterative application of join operations in numerical abstract domains in order to cover positive states without including negative ones.  Our method is tightly integrated to the ICE-DT schema, leading to an efficient data-driven invariant synthesis and verification algorithm. 

Future work includes several directions. First, alternative methods for constructing separators should be investigated in order to reduce the size of the pool of attributes along the learning process while increasing their potential relevance. Another issue to investigate is the control of the counterexamples provided by the teacher since they play an important role in the learning process. In our current implementation, their choice is totally dependent on the SMT solver used for implementing the teacher. Finally, we intend to extend this approach to other types of programs, in particular to programs with other data types, and programs with more general control structures such as procedural programs. 
\bibliographystyle{splncs04} 
\bibliography{bibliography}

\end{document}